\documentclass{article}

\usepackage{spconf,amsmath,graphicx}
\usepackage{amssymb}
\usepackage{enumitem}
\usepackage{hyperref}
\usepackage{amsmath}
\usepackage{comment}
\usepackage{amsmath,amssymb,epsfig}
\usepackage{bm,cite}
\usepackage{verbatim}
\usepackage{subcaption}
\usepackage{setspace}
\usepackage[x11names]{xcolor}
\usepackage[belowskip=2pt,aboveskip=10pt]{caption}
\usepackage{footmisc}
\usepackage{xcolor}
\usepackage[ruled]{algorithm2e}

\ninept
\DeclareMathOperator*{\minA}{min}
\newcommand{\bbR}{\ensuremath{\mathbb{R}}}
\newcommand{\bbE}{\ensuremath{\mathbb{E}}}
\newcommand{\bbN}{\ensuremath{\mathbb{N}}}
\newcommand{\bbC}{\ensuremath{\mathbb{C}}}
\newcommand{\cW}{\ensuremath{\mathcal{W}}}
\newcommand{\cT}{\ensuremath{\mathcal{T}}}
\newcommand{\cA}{\ensuremath{\mathcal{A}}}
\newcommand{\ba}{\ensuremath{\mathbf{a}}}

\newcommand{\bg}{\ensuremath{\mathbf{g}}}
\newcommand{\bmm}{\ensuremath{\mathbf{m}}}
\newcommand{\bh}{\ensuremath{\mathbf{h}}}

\newcommand{\bw}{\ensuremath{\mathbf{w}}}
\newcommand{\bq}{\ensuremath{\mathbf{q}}}
\newcommand{\bs}{\ensuremath{\mathbf{s}}}
\newcommand{\by}{\ensuremath{\mathbf{y}}}
\newcommand{\br}{\ensuremath{\mathbf{r}}}
\newcommand{\bk}{\ensuremath{\mathbf{k}}}
\newcommand{\bc}{\ensuremath{\mathbf{c}}}
\newcommand{\bp}{\ensuremath{\mathbf{p}}}
\newcommand{\bd}{\ensuremath{\mathbf{d}}}
\newcommand{\bz}{\ensuremath{\mathbf{z}}}
\newcommand{\balpha}{\ensuremath{\boldsymbol{\alpha}}}
\newcommand{\btheta}{\ensuremath{\boldsymbol{\theta}}}

\newcommand{\btau}{\ensuremath{\boldsymbol{\tau}}}
\newcommand{\bgama}{\ensuremath{\boldsymbol{\gamma}}}

\newcommand{\bGama}{\ensuremath{\mathbf{\Gamma}}}
\newcommand{\bW}{\ensuremath{\mathbf{W}}}
\newcommand{\bX}{\ensuremath{\mathbf{X}}}

\newcommand{\bM}{\ensuremath{\mathbf{M}}}
\newcommand{\bC}{\ensuremath{\mathbf{C}}}
\newcommand{\bB}{\ensuremath{\mathbf{B}}}

\newcommand{\bA}{\ensuremath{\mathbf{A}}}
\newcommand{\bR}{\ensuremath{\mathbf{R}}}
\newcommand{\bZ}{\ensuremath{\mathbf{Z}}}
\newcommand{\bSig}{\ensuremath{\mathbf{\Sigma}}}
\newcommand{\bsig}{\ensuremath{\boldsymbol{\sigma}}}
\newcommand{\bD}{\ensuremath{\mathbf{D}}}

\newcommand{\bI}{\ensuremath{\mathbf{I}}}

\newcommand{\bK}{\ensuremath{\mathbf{K}}}
\newcommand{\bQ}{\ensuremath{\mathbf{Q}}}
\newcommand{\bL}{\ensuremath{\mathbf{L}}}

\newif\ifproofread

    \title{Joint blind calibration and time-delay estimation for multiband ranging \footnotemark}

    \name{Tarik Kazaz, Mario Coutino, Gerard J. M. Janssen and Alle-Jan van der Veen \thanks{This research was supported in part by NWO-STW under contract 13970 (``SuperGPS''). Mario Coutino is partially supported by CONACYT.}}
    \address{Faculty of EEMCS, Delft University of Technology, Delft, The Netherlands}

\begin{document}
\proofreadfalse
\maketitle


\begin{abstract}
In this paper, we focus on the problem of blind joint calibration of multiband transceivers and time-delay (TD) estimation of multipath channels. We show that this problem can be formulated as a particular case of covariance matching. Although this problem is severely ill-posed, prior information about radio-frequency chain distortions and multipath channel sparsity is used for regularization. This approach leads to a biconvex optimization problem, which is formulated as a rank-constrained linear system and solved by a simple group Lasso algorithm. 
Numerical experiments show that the proposed algorithm provides better calibration and higher resolution for TD estimation than current state-of-the-art methods.
\end{abstract}


\section{Introduction}
Localization in communication networks often requires the estimation of the range between sensor and anchor nodes \cite{gezici2005localization}. The ranging typically starts with the exchange of known probing signals and the estimation of the channel, i.e., the RF chains effect and multipath propagation, between nodes \cite{kazaz2018joint}. 
While for communication the channel is estimated to perform equalization, in localization scenarios, the goal is to remove any bias in range estimation introduced by it.

As a large frequency band (aperture) must be covered during channel probing to increase the resolution of range estimates \cite{witrisal2009noncoherent}, nodes are required to have integrated wideband RF chains. These often introduce frequency-dependent gain and phase distortions in the probing signals due to the used amplifiers and anti-aliasing filters \cite{bazzi2017blind}. For instance, consider low-rate acquisition of multipath signals \cite{gedalyahu2010time, vetterli2001sampling, eldar2015sampling}. Due to the large frequency aperture required during sampling~\cite{kazaz2019multiresolution}, sampling methods are impaired by distortions introduced in RF chains. As these effects can significantly deteriorate range estimation, they need to be estimated and corrected in a \emph{calibration} process. 
Unfortunately, in localization scenarios, calibration is challenging as the sensor nodes are diverse, and manual calibration of each node is not practical.



As calibration is common in many fields, e.g., communications \cite{weiss1990eigenstructure, paulraj1985direction,han2015calibrating}, radio astronomy \cite{wijnholds2009calibration, wijnholds2009multisource, van2019signal}, and medical imaging \cite{van2018calibration}, many algorithms have been proposed for blind calibration. While some of them assume prior knowledge of the measurement matrix, e.g., array response, or the second-order statistics of the calibration parameters, others rely on the Toeplitz structure of the covariance matrix related to the underlying sensor array. Differently from these works, we exploit the properties of the communication channel and formulate the joint blind calibration and time-delay estimation as a special case of a covariance matching problem \cite{ottersten1998covariance}. Even though this formulation leads to an ill-posed problem, using prior information about the distortions of RF chains and the sparsity of the multipath channel the problem can be regularized. Here, we consider that gain distortions of RF chains are slowly varying with frequency, while phase distortions are negligible~\cite{xie2012single}. This assumption allows us to approximate the distortions of the RF chains with a set of known basis functions, leading to a biconvex problem in the calibration and time-delay (TD) parameters. Although biconvex optimization algorithms are applicable, e.g., \cite{ling2015self, friedlander2014bilinear}, the approach in~\cite{ling2015self} does not consider multiple measurement scenarios, and the algorithm in~\cite{friedlander2014bilinear} has a high latency and no convergence guarantees. Therefore, we propose to re-cast the biconvex optimization problem as a rank-1 constrained linear system using the \emph{lifting} technique~\cite{candes2013phaselift, ahmed2013blind, davenport2016overview}, which can be solved efficiently as a group Lasso problem. The proposed algorithm is benchmarked through simulations by comparing its performance with algorithms proposed in \cite{friedlander2014bilinear, ramamohan2019blind}. The results show that the proposed algorithm provides better calibration performance and a higher resolution for TD estimation.

\section{Problem Formulation and Data Model}
 Consider an ultra-wideband (UWB) channel model defined by its continuous-time impulse and frequency response as
\begin{equation}
\label{eq:chan_imp}
	h(t) = \sum_{k=1}^{K} \alpha_k \delta(t-\tau_k) 
	\quad \text{and} \quad 
	H(\omega) = \sum_{k=1}^{K} \alpha_k e^{-j\omega\tau_k},
\end{equation}
\noindent where $K$ is the number of resolvable multipath components (MPCs), $\balpha = [\alpha_1,\dots,\alpha_K]^T \in \bbC^K$ and $\btau = [\tau_1,\dots,\tau_K]^T \in \bbR_+^K$ collect unknown gains and TDs of the MPCs, respectively \cite{molisch2006comprehensive}. We assume that the gains of the MPCs are slowly varying with time according to a Rician distribution. 

In this work, we are interested in estimating $\balpha$  and $\btau$ by probing the channel using the known wideband OFDM probing signal $s(t)$ transmitted over $i = 0, \dots, L-1$, frequency bands. The probed frequency bands are $\cW_i = [\omega_i-\frac{B}{2}, \omega_i + \frac{B}{2}]$, where $B$ is the bandwidth, and $\omega_i$ is the central angular frequency of the $i$th band. The channel probing is performed $P$ times during the channel coherence time. During this time, we assume that multipath gains are slowly varying. We consider that realistic transceivers are used for the channel probing. Our objective then is to perform blind calibration of RF chains and estimate the TDs, $\btau$, from the collected measurements. 

\noindent\textbf{Continous-time signal model:} We consider a baseband signal model and assume ideal conversion to and from the passband. The unknown response of the RF chains at the $i$th band is modeled using equivalent linear and time-invariant low-pass filters ${g_i(t) = g_{{\rm Tx},i}(t) \ast g_{{\rm Rx},i}(t)}$, where the corresponding CTFT $G_i(\omega) = G_{{\rm Tx},i}(\omega) G_{{\rm Rx},i}(\omega)$ has passband $[-\frac{B}{2}, \frac{B}{2}]$. The compound impulse response of the RF chains and the channel is $c_i(t)  = g_i(t) \ast h_i(t)$, where $h_i(t)$ is the baseband equivalent impulse response of the $i$th channel band, and its CTFT is $H_i(\omega) = H(\omega_i+\omega)$.

Consider that there is no inter-symbol interference and that OFDM probing signal is defined as
\begin{equation}
    \nonumber
    \label{eq:pilot_sig}
       s(t)  =
       \begin{cases}
       \sum_{n=0}^{N-1} s_n e^{j \omega_{sc}nt}, & t \in [-T_{\rm CP}, T_{\rm sym}]
       \\
       \mbox{0}, & \text{otherwise}\,,
       \end{cases}
\end{equation}
where the known pilot symbols are $\bs = [s_0,\dots,s_{N-1}]^T \in \bbC^N$, the sub-carrier spacing is $\omega_{\rm sc}$, $T_{\rm CP}$ is the duration of the cyclic prefix, and \(T_{\rm sym} = 2\pi/\omega_{\rm sc}\) is the duration of one symbol. The CTFT of the signal received at the $i$th band after conversion to the baseband and low-pass filtering is 
\begin{equation}
    \label{eq:freq_rx_sig}
       Y_i(\omega)  = S(\omega) C_i(\omega)  + W_i(\omega), \quad \omega \in [-\frac{B}{2}, \frac{B}{2}]\,,
\end{equation}
\noindent where $Y_i(\omega) = 0$, otherwise, $C_i(\omega) = G_i(\omega) H_i(\omega)$, and $W_i(\omega)$ is low-pass filtered Gaussian white noise.

\noindent\textbf{Discrete-time signal model:} The receiver samples $y_i(t)$ with period $T_s=1/B$, performs packet detection, symbol synchronization, and removes the cyclic prefix. During the duration of one symbol $N$ complex samples are collected, i.e., $T_{\rm sym} = NT_s$. Next, a $N$-point DFT is applied on the collected samples, and DFT coefficients obtained during the $p$th probing interval are stacked in the increasing order of the their frequencies in $\by_i(p) \in \bbC^{N}$. The discrete data model of the processed signals [cf.~\eqref{eq:freq_rx_sig}] received during $P$ probing intervals can be written as
\begin{equation}
    \label{eq:data_model}
        \by_i(p) = \text{diag}(\bs) \bc_i(p) + \bw_i(p)\,, \quad p = 1, \dots, P\,,
\end{equation}
\noindent where $\bc_i = \text{diag}(\bg_i) \bh_i(p)$, and $\bw_i(p) \in \bbC^{N}$ is zero-mean white Gaussian distributed noise. The samples of $G_i(\omega)$ at the subcarrier frequencies are collected in $\bg_i = [g_{i,0}, \dots,g_{i,N-1}]^T \in \bbC^{N}$, where $g_{i,n} = \rho_{i,n}e^{j\psi_{i,n}}$ with $\rho_{i,n}$ and $\psi_{i,n}$ denoting the unknown gain and phase distortions of the RF chains, respectively. Likewise, $\bh_i(p) \in \bbC^{N}$ collects samples of $H_i(\omega)$ in increasing order of frequencies as 
\begin{equation}
    \label{eq:ch_dis1}
        H_i[n] = H\left(\omega_i + n\omega_{\rm sc} \right), \quad n = -\frac{N}{2}, \dots, \frac{N}{2}\,,
\end{equation}
\noindent where $\omega_{\rm sc} = \frac{2\pi}{NT_s}$, and we assume that $N$ is an even number. We consider that bands $\left\{\cW_i\right\}_{i=0}^{L-1}$ are laying on the discrete frequency grid $\omega_i = \omega_0 + n_i\omega_{\rm sc}$, where $n_i \in \bbN$, and $\omega_{0}$ denotes the lowest frequency considered during channel probing. 

Inserting the channel model (\ref{eq:chan_imp}) into (\ref{eq:ch_dis1}) gives
\begin{equation}
    \label{eq:ch_dis2}
        H_i[n] = \sum_{k=1}^{K} \alpha_k e^{-jn_i\omega_{\rm sc}\tau_k}e^{-jn\omega_{\rm sc}\tau_k},
\end{equation}
where we absorbed $e^{-j\omega_{0}\tau_k}$ in $\alpha_k\,\forall\,k$. The channel vector $\bh_i(p)$ satisfies the model
\begin{equation}
    \label{eq:ch_model}
        \bh_i(p) = \bM\text{diag}(\btheta_i)\balpha(p)\,,
\end{equation} 
where  $\bM = [\bmm_1, \dots, \bmm_K] \in \bbC^{N\times K}$ is a Vandermonde matrix with its $k$th column given by
\begin{equation}
\nonumber
\label{eq:vandermonde1}
\bmm_k = \left[1, e^{-j\omega_{\rm sc}\tau_k}, \dots, e^{-j(N-1)\omega_{\rm sc}\tau_k}\right]^T\,.
\end{equation}
\noindent Likewise, $\btheta_i = [\theta_{i,1},\dots,\theta_{i,K}]^T \in \bbC^K$, where $\theta_{i,k} = e^{-jn_i\omega_{\rm sc}\tau_k}$ are band-dependent phase shifts of the MPCs.

\noindent \textbf{Data model:} Each $\bc_i$ is estimated by deconvolution of (\ref{eq:data_model}) as 
\begin{equation}
\nonumber
    \label{eq:ch_est}
        \bc_i(p) = \text{diag}^{-1}(\bs)\by_i(p)\,.
\end{equation}
\noindent The deconvolved measurements satisfy the model 
\begin{equation}
    \label{eq:ch_est_model}
        \bc_i(p) = \text{diag}(\bg_i)\bM\text{diag}(\btheta_i)\balpha(p) + \bw_i'(p)\,,
\end{equation}
\noindent where the pilot symbols have constant magnitude and $\bw_i'(p) = \text{diag}^{-1}(\bs)\bw_i(p)$ is zero-mean white Gaussian distributed noise.

The estimates of the compound frequency response, $\bc_i(p)$, are stacked in $\bc(p) = [\bc_1^T(p),\dots,\bc_L^T(p)]^T \in \bbC^{NL}$. From (\ref{eq:ch_est_model}), the model for $\bc(p)$ is
\begin{equation}
    \label{eq:ch_est_model2}
        \bc(p) = \text{diag}(\bg)\bA(\btau)\balpha(p) + \bw(p)\,,
\end{equation}

\noindent where 
$\bA(\btau) = [\ba(\tau_1),\dots,\ba(\tau_K)] \in \bbC^{NL \times K}$ has the multiple invariance structure
\begin{equation}
\nonumber
\label{eq:channel_vector}
  \bA(\btau) =
    \begin{bmatrix} 
      \bM \\
      \bM\text{diag}(\btheta_1)\\
      \vdots\\
      \bM\text{diag}(\btheta_{L-1})\\
  \end{bmatrix}\,,\qquad
    \bg = \begin{bmatrix}
      \bg_{1} \\ \bg_{2} \\ \vdots \\ \bg_{L}
    \end{bmatrix} \,,
\end{equation}
\noindent and likewise,  $\bw(p) \in \bbC^{NL}$ collects $\bw_i'(p)$, $i=0,\dots,L-1$.
\noindent

Stacking all the estimates of the compound frequency responses, collected during $P$ probing intervals in $\bC =\left[\bc(1),\dots,\bc(P)\right] \in\bbC^{NL \times P}$, leads to the model [cf.\eqref{eq:ch_est_model2}]
\begin{equation}
    \label{eq:channel_matrix}
        \bC = \text{diag}(\bg)\bA(\btau)\bX + \bW\,,
\end{equation}
\noindent where $\bX = \left[\balpha(1),\dots,\balpha(P)\right] \in \bbC^{K \times P}$, and $\bW$ collects $\bw(p)$ $\forall\; p$.

\section{Joint blind calibration and TD estimation}
Our objective is to estimate the unknown response of the RF chains, $\bg$, and TDs, $\btau$, of the MPCs from the measurement matrix $\bC$. We first introduce a general problem, and then propose an efficient algorithm for solving it. Joint blind calibration and TD estimation can be formulated as the following optimization problem
\begin{equation}
    \label{eq:cs_opt1}
        \hat{\bg}, \hat{\btau}, \hat{\bX} = \minA_{\bg, \btau, \bX} \Vert \bC - \text{diag}(\bg)\bA(\btau)\bX \Vert_{F}^2\,,
\end{equation}
\noindent where $\Vert \cdot \Vert_{F}$ is the Frobenius norm of a matrix. This problem is clearly ill-posed and non linear, making it difficult to solve without further assumptions or prior information. Therefore, we use prior knowledge about the frequency response of RF chains and the sparsity of the multipath channels to reformulate the problem.

\noindent \textbf{Assumptions:} The magnitude of the frequency response of RF chains is slowly varying with frequency, while phase distortions are usually negligible \cite{xie2012single}. Therefore, the entries of $\bg$ are slowly changing, and it can be approximated as $\bg = \bB\bp$, where the columns of $\bB \in \bbC^{NL \times R}$ are $R$ known basis functions and $\bp$ are unknown calibrating parameters. In this paper, we assume that columns of $\bB$ are the first $R$ Chebyshev polynomials of the first kind as they offer near minimax polynomial approximation of $\bg$ in an interval \cite{bistritz1979model}. 

Let the maximum expected TD to be estimated in the ranging scenario be $\tau_{\rm max} = \frac{d_{\rm max}}{c} + \tau_{\rm tot}$, where $d_{\rm max}$ is the maximum distance, $c$ is the speed of light, and $\tau_{\rm tot}$ is the total delay spread of the channel. Assuming that the unknown TDs lay on a uniform grid of $M \gg NL$ delays, i.e., $\tau_k \in \cT = \{ 0, \frac{\tau_{\rm max}}{M}, \dots, \frac{\tau_{\rm max}(M-1)}{M} \}$, the following optimization problem can be formulated to solve the joint blind calibration and TD estimation
\begin{equation}
    \label{eq:cs_opt2}
        \hat{\bp}, \hat{\bX}_s = \minA_{\bp, \bX_s} \Vert \bC - \text{diag}(\bB\bp)\bA_D\bX_s \Vert_{F}^2 + \lambda \Vert \bX_s^T \Vert_{2,1}\,,
\end{equation}
\noindent where $\bA_D = [\ba(t_0),\dots,\ba(t_{M-1})] \in \bbC^{NL \times M}$ is  a dictionary matrix with column vector defined in (\ref{eq:ch_est_model2}), $t_m = \frac{m}{M}\tau_{\rm max}$ and $\bX_s \in \bbC^{M \times P}$ is a \emph{row sparse matrix}. The regularization parameter $\lambda > 0$ is  determining the sparsity (i.e., number of non-zero rows in $\bX_s$), and $\Vert \bm [\bm a_1,\ldots,\bm a_n] \Vert_{2,1} := \sum_{i=1}^{n}\Vert \bm a_i \Vert_2$ is the $\ell_{2,1}$-norm of a matrix which is known to promote column sparsity.

Although the optimization problem in (\ref{eq:cs_opt2}) is biconvex, i.e. it is convex in $\bp$ for fixed $\bX_s$ and convex in $\bX_s$ for fixed $\bp$, and alternating minimization can be used to estimate both $\bX_s$ and $\bp$, the lack of convergence guarantees and the high computational complexity makes~\eqref{eq:cs_opt2} unpractical. Therefore, we propose a method that offers a better solution using ideas of covariance matching.

Let us assume that (\ref{eq:chan_imp}) is a wide-sense stationary and uncorrelated scattering (WSSUS) fading channel. Therefore, $\balpha(p)$ and $\bw(p)$ are statistically independent and mutually uncorrelated variables with covariance matrices $\bSig_{\alpha}=\text{diag}(\bsig_{\alpha})$, $\bsig_{\alpha} = [\sigma_{\alpha,1}^2,\dots,\sigma_{\alpha,K}^2]^T$, and $\bSig_{w}=\sigma_w^2\bI_{NL}$, where $\bI_{NL}$ is the $NL \times NL$ identity matrix \cite{pedersen2011analysis}. With these assumptions, we can write the covariance matrix of $\bc(p)$ as
\begin{equation}
    \label{eq:cov_model1}
    \begin{split}
        \bR_{c} & := \bbE\{\bc(p)\bc^H(p)\} \in \bbC^{NL \times NL}\,,
        \\ & =\text{diag}(\bg)\bA(\btau)\bSig_{\alpha}\bA^H(\btau)\text{diag}(\bar{\bg}) + \sigma_w^2\bI_{NL}\,,
    \end{split}
\end{equation}
\noindent where $\overline{(\cdot)}$ denotes complex conjugation. To obtain a linear
measurement model, we vectorize (\ref{eq:cov_model1}) and write it as
\begin{equation}
    \label{eq:cov_model2}
    \br_{c} = \text{diag}(\overline{\bg} \otimes \bg) \bK(\btau) \bsig_{\alpha} + \br_w \,,
\end{equation}
where $\otimes$ is the Kronecker product, $\bK(\btau) = \overline{\bA}(\btau)\circ \bA(\btau) \in \bbC^{(NL)^2 \times K}$, $\circ$ denotes the Khatri-Rao product and $\br_w = \sigma_w^2 \text{vec}(\bI_{NL})$. Here, $\text{vec}(\cdot)$ stacks the columns of the matrix. 

\noindent\textbf{Algorithm:} The covariance matrix can be estimated from measurements as $\hat{\bR}_{c} = \frac{1}{P}\bC\bC^H$, where its vectorized form is $\hat{\br}_{c} = \text{vec}(\hat{\bR}_{c})$. Here, we assume \textit{a priori} knowledge of the noise power $\sigma_w^2$, and we define $\tilde{\br}_{c} = \hat{\br}_{c} - \br_w$. For the case with unknown $\sigma_w^2$, we can first estimate it according to \cite{bresler1988maximum}. Considering the modelling assumption on $\bg$ and multipath channel sparsity, and using the properties of the Kronecker product, we can rewrite (\ref{eq:cov_model2}) as
\begin{equation}
    \label{eq:cov_model4}
    \br_{c} = \text{diag}(\bD \bz) \bK_D \br_{\alpha} + \br_w\,,
\end{equation}
\noindent where $\bD = \overline{\bB} \otimes \bB$ has size $(NL)^2\times R^2$, $\bz = \overline{\bp} \otimes \bp \in \bbC^{R^2}$, $\bK_D = \overline{\bA}_D\circ \bA_D\in \bbC^{(NL)^2 \times M}$ is a dictionary matrix and $\br_{\alpha} \in \bbR^M$ is a $K$ sparse vector that collects the powers of the MPCs. The unknown parameters in the data model are the calibrating parameters $\bz$ and the powers of the MPCs $\br_{\alpha}$. Note that finding the columns of $\bK_D$ that correspond to the non-zero elements of $\br_{\alpha}$ is equivalent to estimating $\btau$. To estimate these parameters, we formulate the following sparse covariance matching optimization problem 
\begin{equation}
    \label{eq:sparse_covm1}
        \hat{\bz}, \hat{\br}_{\alpha} = \minA_{\bz, \br_{\alpha}} \Vert \tilde{\br}_{c} -  \text{diag}(\bD \bz) \bK_D \br_{\alpha} \Vert_{2}^2 + \lambda \Vert \br_{\alpha} \Vert_{1}\,,
\end{equation}
\noindent where $\Vert \cdot \Vert_{2}$ denotes the $\ell_{2}$-norm of the vector, $\lambda > 0$ controls the level of sparsity of $\br_{\alpha}$, and $\Vert \cdot \Vert_{1}$ is the $\ell_{1}$-norm of a vector.

Similar to (\ref{eq:cs_opt2}), the objective function of this optimization problem is biconvex in the unknown parameters $\bz$ and $\br_{\alpha}$. To alleviate difficulties arising from the biconvexity of the objective function, we reformulate (\ref{eq:sparse_covm1}) as a problem involving solving a linear system whose solution obeys a rank-1 constraint by \emph{lifting} the unknown variables. The elements of $\br_{c}$ can be written as
\begin{equation}
    \label{eq:lifting1}
    [\br_{c}]_n = [\bD \bz]_n \bk_n^T \br_{\alpha} + [\br_w]_n = \bd_n^T \bz \br_{\alpha}^T \bk_n + [\br_w]_n\,,\; \forall\,n,
\end{equation}
where $\bd_n^T$ and $\bk_n^T$ denote the $n$th row of $\bD$ and $\bK_D$, respectively. Let us define the rank-1 matrix $\bQ := \bz\br_{\alpha}^T$ and the linear operator $\cA : \bbC^{R^2 \times M} \to \bbC^{NL}$ as
\begin{equation}
    \label{eq:lifting2}
    \br_{c} = \cA(\bQ) + \br_w := \text{vec}(\{ \bd_n^T \bQ \bk_n \}_{n=1}^{NL}) + \br_w\,.
\end{equation}
Given that $\bd_n^T \bQ \bk_n = (\bk_n \otimes \bd_n)^T\text{vec}(\bQ)\,\forall\, n$, (\ref{eq:lifting2}) becomes
\begin{equation}
    \label{eq:matrix_form}
    \br_{c} = \bGama \bq + \br_w\,,
\end{equation}
\noindent where the $n$th row of $\bGama \in \bbC^{NL \times R^2 M}$ (the matrix representation of the operator $\cA$) is $\bgama_n^T = (\bk_n \otimes \bd_n)^T$, and $\bq = \text{vec}(\bQ)$.

The problem of estimating $\bz$ and $\br_{\alpha}$ then reduces to finding a rank-1 matrix $\bQ$ satisfying the set of linear constrains (\ref{eq:matrix_form}). The solution of this problem can be found by 
\begin{equation}
    \label{eq:nuc_problem}
        \hat{\bQ} = \minA_{\bQ} \Vert \tilde{\br}_{c} - \bGama \text{vec(\bQ)} \Vert_{2}^2 + \lambda \Vert \bQ \Vert_{*}\,,
\end{equation}
\noindent where $\Vert \cdot \Vert_{*}$ denotes the nuclear norm of a matrix which promotes low rank solutions. To further simplify the problem, we use the sparsity of $\bQ$. Due to $\br_{\alpha}$, the matrix $\bQ$ is not only rank-1 but also column sparse. Since for any matrix $\bL$, $\Vert \bL \Vert_{2,1} > \Vert \bL \Vert_{*}$ holds, we can use the $\Vert \cdot \Vert_{2,1}$-norm to regularize (\ref{eq:nuc_problem}) instead of $\Vert\cdot\Vert_*$ following~\cite{ling2015self} and obtain a simpler formulation. Therefore, to estimate $\bz$ and $\br_{\alpha}$ it is sufficient to solve
\begin{equation}
    \label{eq:l1_problem}
        \hat{\bq} = \underset{\bq}{\arg\minA} \Vert \tilde{\br}_{c} - \bGama \text{vec}(\bQ) \Vert_{2}^2 + \lambda \Vert \bQ \Vert_{2,1}\,,
\end{equation}
\noindent where the regularization parameter $\lambda > 0$ is set to be proportional to the noise power $\sigma_w^2$. This problem, besides of being convex, can be identified as a \emph{group Lasso} problem, which can be solved efficiently. Here, we use the spectral gradient-projection method \cite{BergFriedlander:2008, spgl1:2007}. 

To estimate $\bz$ and $\bm \sigma_{\alpha}$ after solving (\ref{eq:l1_problem}), first $\bQ$ is reconstructed from $\bq$, and then the singular value decomposition is used to find the best rank-1 approximation of $\bQ$ in the $\ell_2$-sense~\cite{van1993subspace}. Then, $\bz$ and $\br_{\alpha}$ are found as the left and right principal singular vectors, respectively. Similarly, to estimate the calibrating parameters $\bp$, first matrix $\bZ \in \bbC^{R \times R}$ is constructed from $\bz$, and then $\bp$ is proportional, up to a complex scaling factor, to the right principal singular vector of $\bZ$. As this scaling ambiguity does not influence performance of the TD estimation, it can be ignored. The estimates for parameters $\bg$, $\btau$, and $\bm\sigma_{\balpha}$ immediately follow. This procedure is summarized in Alg.~\ref{alg:bjctd} \footnote{MATLAB notation has been used for simplicity.}.

\begin{algorithm}[t]
\SetAlgoLined
\KwIn{$\{\cT,\bB,\bGama,\tilde{\br}_{c},\lambda\}$}
 $\hat{\bq} \leftarrow {\arg\minA}_{\bq}\; \Vert \tilde{\br}_{c} - \bGama \text{vec}(\bQ) \Vert_{2}^2 + \lambda \Vert \bQ \Vert_{2,1}$\;
 $\hat{\bQ} \leftarrow {\rm unvec}(\hat{\bq})$\;
 $\{\hat{\bz},\sim,\hat{\br}_s\} \leftarrow\texttt{svds}(\bQ,1)$\;
 $\bZ \leftarrow{\rm unvec}(\hat{\bz})$\;
 $\{\sim,\sim,\hat{\bp}\}\leftarrow{\rm svds}(\bZ,1)$\;
 $\hat{\bg} \leftarrow \bB\hat{\bp}$\;
 $\{\hat{\bm \sigma}_{\balpha},{\rm indxSet}\} \leftarrow$ \texttt{find}($\hat{\br}_s\, \sim= 0$)\;
 $\hat{\btau} \leftarrow{\cT}(\rm{indxSet})$\;
 \KwOut{$\{\hat{\bp},\hat{\bg},\hat{\bm\sigma}_{\balpha},\hat{\btau}\}$}
 \caption{Joint Blind Calibration and TD Estimation}
 \label{alg:bjctd}
\end{algorithm}
\setlength{\textfloatsep}{2pt}
As the resolution of $\btau$ estimates from ($\ref{eq:l1_problem}$) is restricted by the resolution of the chosen grid $\cT$, in case that the TDs $\btau$ do not lie exactly on the grid $\cT$, this algorithm can be extended with grid-less estimation methods such as multiple invariance ESPRIT \cite{kazaz2019sub, miesprit1992}. 

\begin{figure*}[t!]
	\centering
	\begin{subfigure}{0.5\textwidth}
		\includegraphics[trim=2 2 0	1,clip,width=7.0cm]{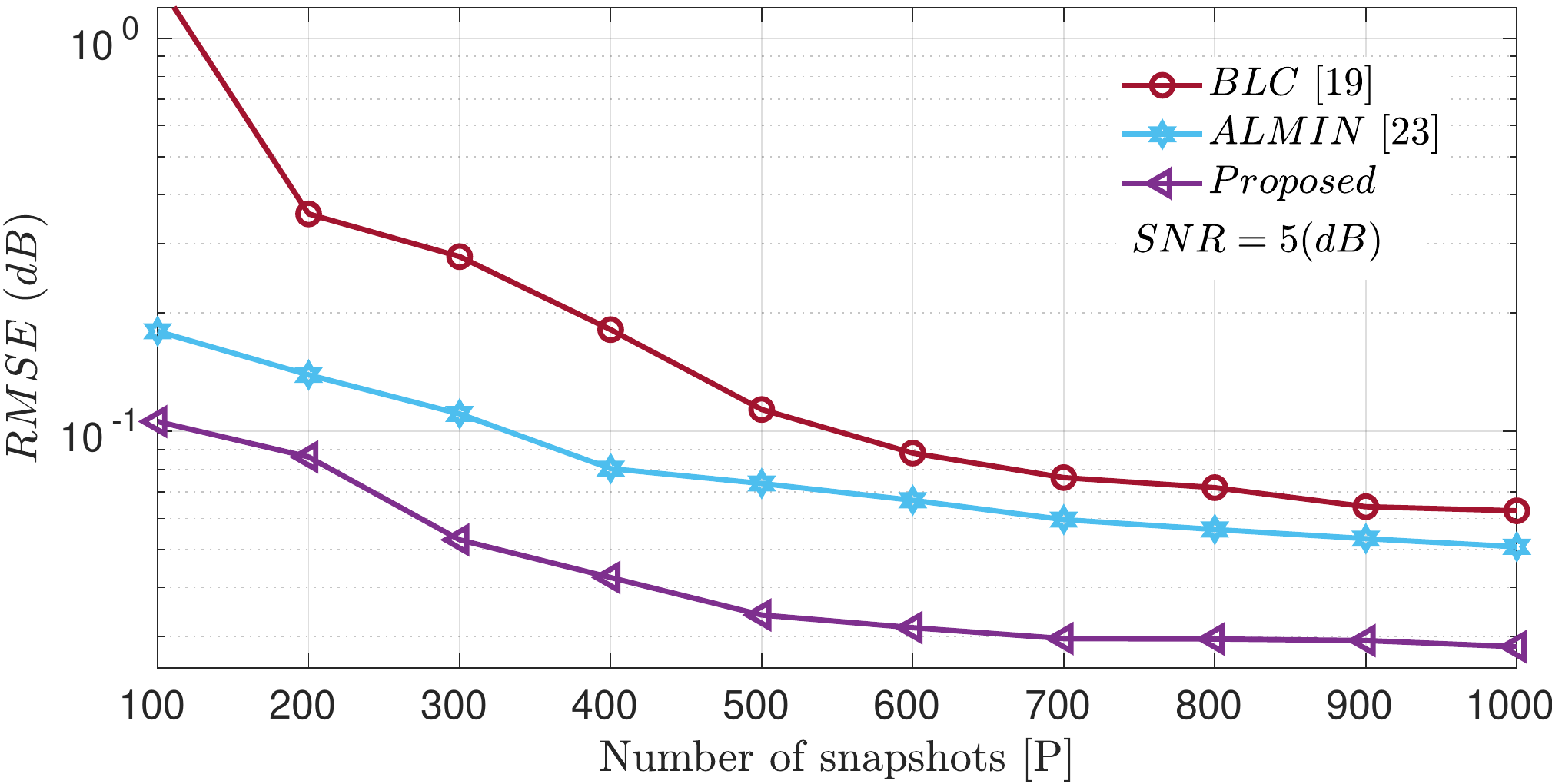}
		\caption{}
	    \label{fig:per:cal_snap}
	\end{subfigure}%
	\begin{subfigure}{0.5\textwidth}
		\includegraphics[trim=1 2 0	1,clip,width=6.9cm]{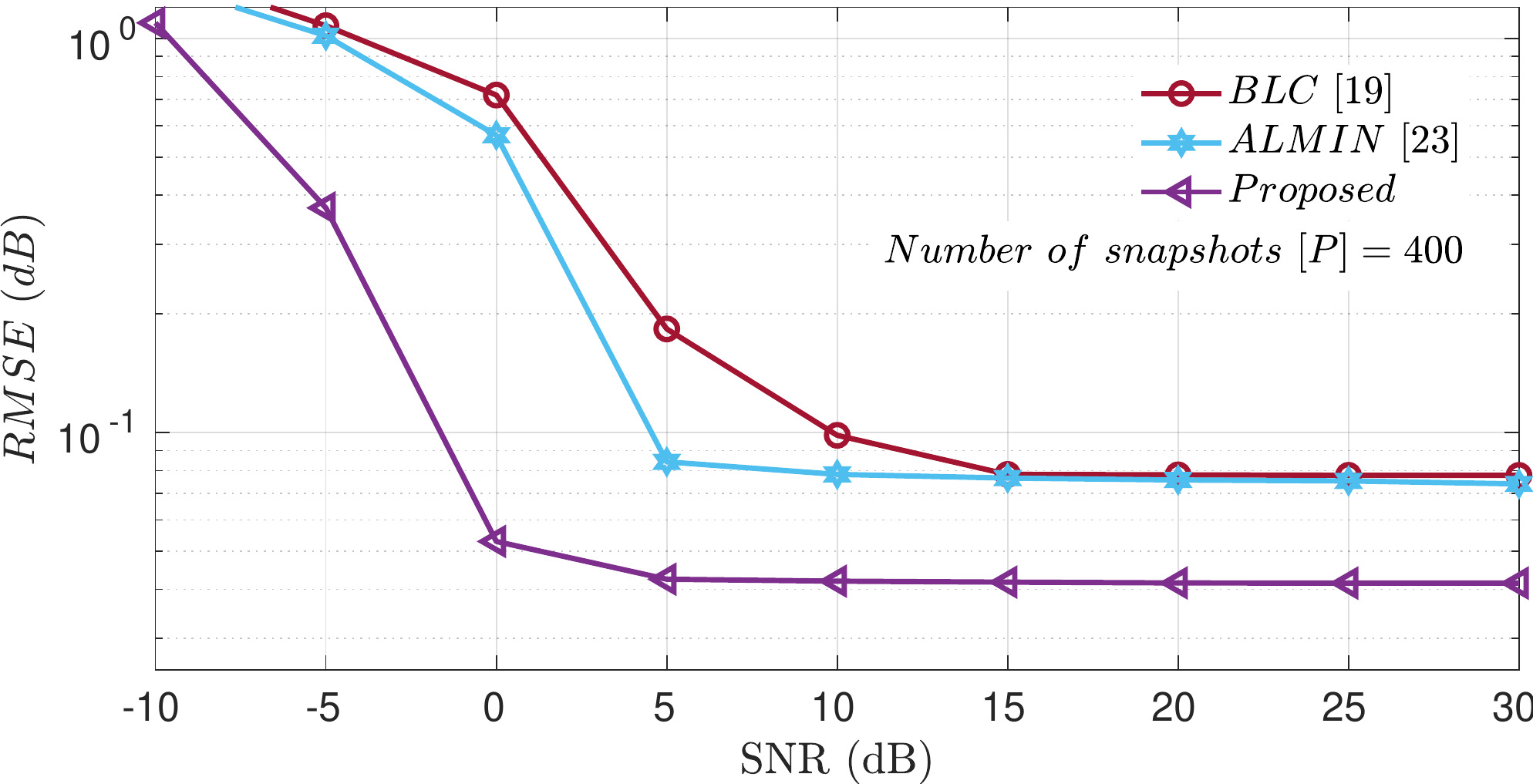}
		\caption{}
	    \label{fig:per:cal_snr}
	\end{subfigure}
	\begin{subfigure}{0.5\textwidth}
	    \hspace{2mm}
		\includegraphics[trim=2 2 0	1,clip,width=6.8cm]{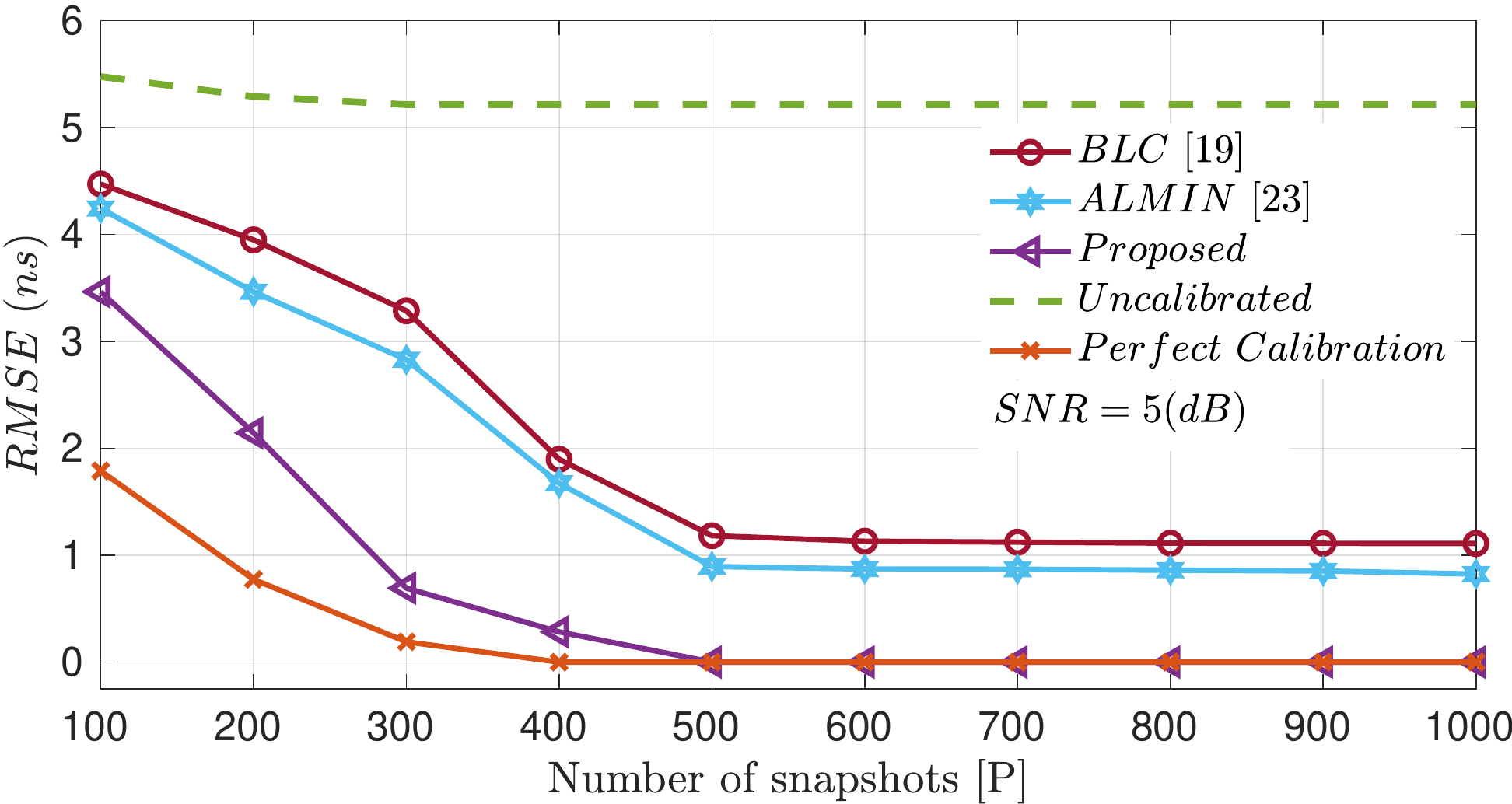}
		\caption{}
	    \label{fig:per:td_snap}
	\end{subfigure}%
	\begin{subfigure}{0.5\textwidth}
	    \hspace{2mm}
		\includegraphics[trim=2 2 0	1,clip,width=6.8cm]{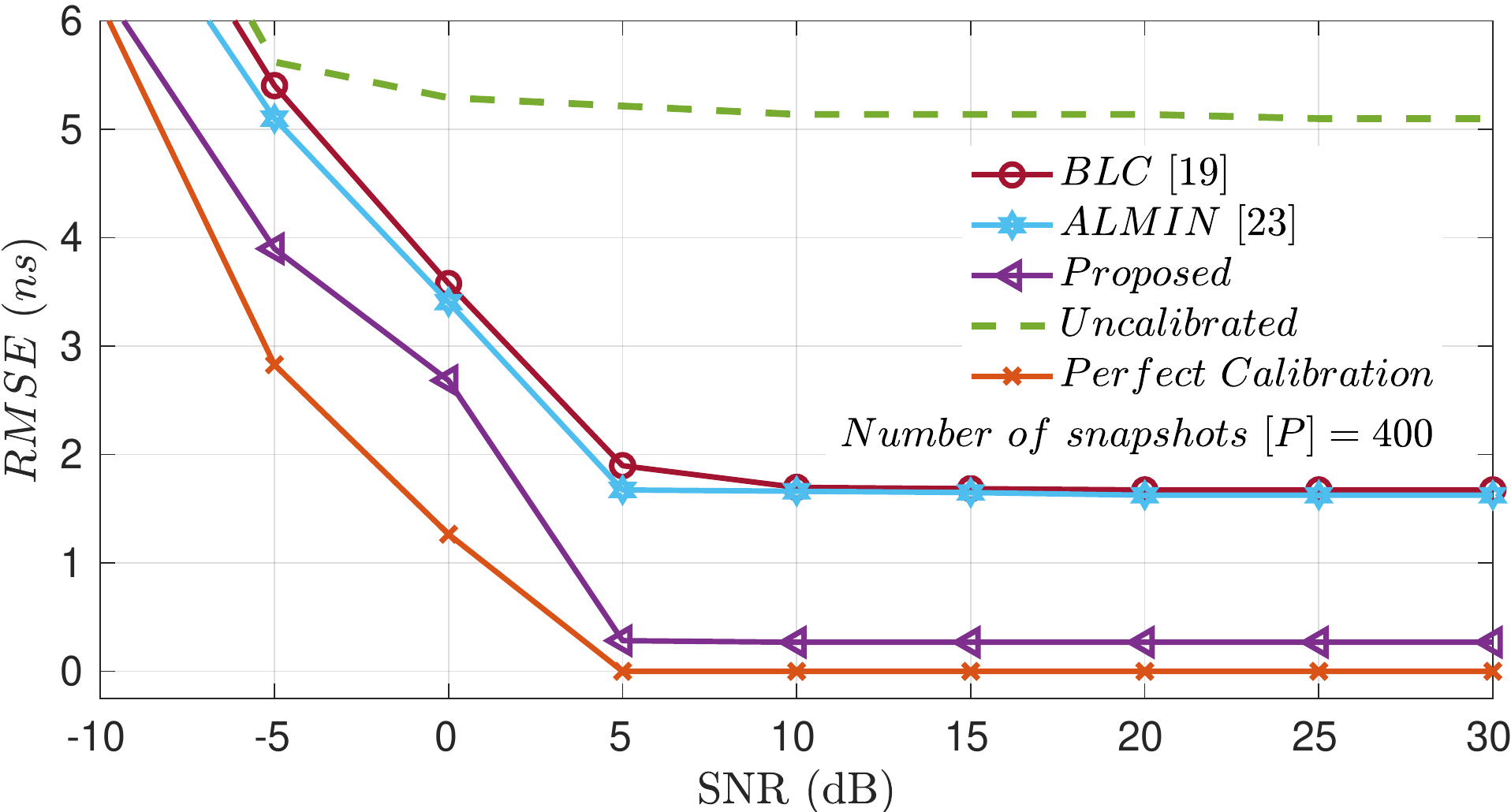}
		\caption{}
	    \label{fig:per:td_snr}
	\end{subfigure}
	\caption{RMSE for estimated calibration (a-b) and time-delay (c-d), vs number of snapshots and signal to noise ratio.}
	\label{fig:per}
\end{figure*}
\section{Numerical Experiments}
This section evaluates the performance of the proposed algorithm via numerical simulations. 
We consider a scenario where the multipath channel has eight dominant MPCs, i.e. $K = 8$, with gains distributed according to a Rician distribution.  The continuous-time channel is modeled using a $2$ GHz grid, with channel tap delays spaced at $500$ ps. 
We consider that the receiver estimates the channel frequency response in four frequency bands, i.e., $L = 4$, using a probing signals with $N = 64$ subcarriers and a bandwidth of $B = 20$ MHz. The central frequencies of the bands are $\{10, 70, 130, 280 \}$ MHz, respectively. The gain errors, i.e., elements of $\bg$, are drawn uniformly from the interval of $[-3, 3]$dB, considering that gain variations are smooth over subcarriers. During the simulations, $\bg$ is kept fixed.  To evaluate performance for TD estimation, we use the root mean Square error (RMSE) of the first multipath component TD estimate. To evaluate the performance of the calibration, we use the average RMSE of the gain estimates over all the subcarriers and bands. The RMSE are computed using $10^3$ independent Monte-Carlo trials and compared  with RMSEs of the algorithms  proposed  in  \cite{friedlander2014bilinear, ramamohan2019blind} which are shortly denoted with ALMIN and BLC, respectively. 

The original formulation of the BLC algorithm does not require knowledge of the noise covariance $\bSig_w$, as the authors in \cite{ ramamohan2019blind} assume that the nonideal response of the sensor array is affecting both signal and noise. While this is typically the case for the acoustic sensor vectors, this assumption does not hold for calibrating RF chains. Therefore, we provide a good initial estimate on the true $\bSig_w$ to the BLC algorithm. Likewise, the ALMIN algorithm is initialized with a good initial guess on $\bg$, and to limit its computational complexity the maximum number of iterations is set to eight. 

Fig.\ \ref{fig:per:cal_snap} shows the calibration performance of the proposed, ALMIN and BLC algorithms with respect to the number of snapshots $P$. The signal to noise ratio (SNR) is set to $5$ dB and kept fixed during trials. From Fig.\ \ref{fig:per:cal_snap}, we observe that calibration RMSE decreases as the number of snapshots increases for all three algorithms due to better estimation of the covariance matrix, $\hat{\bR}_c$, and a better model matching [cf.~\eqref{eq:l1_problem}].

In the second scenario, we fixed the number of snapshots to $P = 400$, and evaluated the methods performance for different SNR regimes. From Fig.\ \ref{fig:per:cal_snr}, it is seen that calibration RMSE decreases with SNR. However, for a SNR above $5$ dB for the proposed and ALMIN algorithm or a SNR above $15$ dB for the BLC algorithm, it saturates, due to model mismatch related to the limited number of snapshots. 

The same simulation scenarios are repeated for the TD estimation, and the corresponding RMSEs are shown in Fig.\ \ref{fig:per:td_snap} and Fig.\ \ref{fig:per:td_snr}. In addition to the algorithms mentioned above, the RMSEs of the estimates obtained using the $\ell_1$ based algorithm with perfectly calibrated and uncalibrated RF chains are shown. From Fig.\ \ref{fig:per:td_snap}, we observe that for a sufficient number of snapshots, the proposed algorithm is able to recover exactly the TD of the first MPC. However, the BLC and ALMIN algorithm are biased due to the model mismatch and the limited number of iterations allowed for convergence, respectively. In Fig.\ \ref{fig:per:td_snr}, the RMSEs are shown for $P=400$ snapshots and different SNR levels. It is seen that in the case of a limited number of snapshots, the RMSEs of all algorithms are saturated for SNR above $5$ dB.  Therefore, all the algorithms are biased compared to the $\ell_1$ based estimation with perfect calibration. This is a consequence of errors in the estimation of the covariance matrix from the limited number of snapshots.

\section{Conclusions} 
In this paper, we proposed an algorithm for joint blind calibration and time-delay estimation for multiband ranging by formulating this problem as a particular case of a covariance matching. Although this problem is severely ill-posed, prior information about RF chain distortions and multipath channel sparsity was used to regularize it. The resulting optimization problem though is biconvex, can be recasted as a rank-1 constrained linear system of equations using \textit{lifting}, which can be solved efficiently using a group Lasso algorithm. Future directions aim towards finding optimal regularization parameters and extensions to support off-grid time-delay estimation. 

\newpage
\clearpage
\bibliographystyle{IEEEtran}
\bibliography{refs}

\end{document}